\documentclass[10pt]{iopart}

\usepackage{iopams,bbm}
\usepackage{setstack}
\usepackage{changes,fixme}
\usepackage{longtable}
\newcommand{\ket}[1]{|#1\rangle}
\newcommand{\bra}[1]{\langle#1|}

\newcommand{\proj}[1]{\ket{#1}\!\bra{#1}}

\usepackage{hyperref}

\begin{document}

\title[Translationally invariant multipartite Bell
inequalities involving $\ldots$]{Translationally invariant
multipartite Bell inequalities involving only two-body correlators}

\author{J Tura$^1$, A B Sainz$^1$, T V\'ertesi$^2$, A Ac\'in$^{1,3}$, M
Lewenstein$^{1,3}$ and R Augusiak$^1$}
\address{$^1$ ICFO--Institut de Ci\`{e}ncies Fot\`{o}niques, 08860
Castelldefels (Barcelona), Spain}
\address{$^2$ Institute for Nuclear Research, Hungarian Academy of
Sciences, H-4001 Debrecen, P.O. Box 51, Hungary}
\address{$^3$ ICREA--Institucio Catalana de Recerca i
Estudis Avan\c{c}ats, Lluis Companys 23, 08010 Barcelona, Spain}
\ead{remigiusz.augusiak@icfo.es}

\begin{abstract}
Bell inequalities are natural tools that allow one to certify the
presence of nonlocality in quantum systems. The known
constructions of multipartite Bell inequalities contain, however,
correlation functions involving all observers, making their
experimental implementation difficult. The main purpose of this
work is to explore the possibility of witnessing nonlocality in
multipartite quantum states from the easiest-to-measure
quantities, that is, the two-body correlations. In particular, we
determine all three and four-partite Bell inequalities constructed
from one and two-body expectation values that obey translational
symmetry, and show that they reveal nonlocality in multipartite
states. Also, by providing a particular example of a five-partite
Bell inequality, we show that nonlocality can be detected from
two-body correlators involving only nearest neighbours. Finally,
we demonstrate that any translationally invariant Bell inequality
can be maximally violated by a translationally invariant state and
the same set of observables at all sites. We provide a numerical
algorithm allowing one to seek for maximal violation of a
translationally invariant Bell inequality.
\end{abstract}

\maketitle

\section{Historical background and introduction}

\textbf{Historical background.}  \textit{The first quantum mechanical
revolution} took place in the beginning of the XXth century. Quantum
mechanics was then discovered and used to describe and explain the
laws of micro-world, and to predict and calculate with great precision
the properties of quantum  systems. The fundamental nature of quantum
mechanics and its paradoxes were, of course, discussed in those times
\cite{ZurekWheeler}, but they were treated as philosophical curiosities,
rather than serious scientific challenges. All of that has changed due
to the seminal work of John Bell \cite{Bellsbook}. As Alain Aspect
termes it, Bell's work has initiated {\it the second quantum
revolution}, that has led to experimental confirmation of non-locality
of quantum mechanics, pioneered by Aspect himself
\cite{Aspectmostquotedpaper}. The same developments triggered
spectacular progress of ultraprecise studies of quantum single and few
particle systems, that culminated with Nobel Prizes of Hans Dehmelt,
Wolfgang Paul and Norman Ramsey in 1989, and Serge Haroche and Dave
Wineland in 2012. Recent developments of quantum information theory
and physics of ultracold matter mark the dawn of {\it the third
quantum mechanical revolution} \cite{Maciekbook}. Quantum mechanics,
after achieving maturity as a physical discipline describing
quantitative and fundamental aspects of the micro-world, becomes
nowadays the base for future quantum technologies.  The paper, that we
present below, belongs to these recent trends, and treats  the
challenging problem of feasible detection of non-locality in many-body
quantum systems.\\

\noindent\textbf{Introduction.} Nonlocality, that is, the
existence of correlations that violate Bell inequalities,
evidences that quantum phenomena cannot be explained by any local
theory (see, e.g., Ref. \cite{BrunnerReview}). Besides this
fundamental significance, nonlocal correlations have also turned
into a key resource for quantum information theory. In particular,
nonlocal correlations perform better than classical resources at
communication complexity problems \cite{CommCompl} (see also Ref.
\cite{CommComplRev}) and enable quantum key distribution
\cite{Ekert91,BHK,Us}. Moreover, they can be used to certify the
presence of true randomness in measurement statistics \cite{Toni},
and also to amplify it
\cite{ColbeckRenner,rodrigo,andrzej,monogamies}, in the sense that
exploiting nonlocal correlations one can obtain perfectly random
bits from partially random ones.

Certification of the presence of nonlocality in quantum states is
therefore one of the central problems in quantum information
theory. The most natural tool that serves the purpose are Bell
inequalities \cite{Bell}. These are linear inequalities,
formulated in terms of expectation values involving products of
local measurements performed by the parties. They are satisfied by
all classical correlations, and possibly violated by some nonlocal
ones. In principle, since classical correlations form a
polytope\footnote{Recall that one defines a polytope to be a
bounded convex set with a finite number of extreme points.},
identifying all Bell inequalities corresponding to its facets
would completely solve the above problem. Considerable effort has
been devoted to achieve this goal and many Bell inequalities have
been found (see, e.g.,
\cite{WernerWolf,ZukowskiBrukner,Wiesiu,Sliwa,JD}). Usually,
however, they contain expectation values involving all observers,
and even if, intuitively, these are the strongest Bell
inequalities, they are hardly applicable in experiments where the
quantum system under consideration is large, in the sense that it
contains a considerable number of particles. In fact, in these
systems the measurement of expectation values involving all the
observers is very challenging.

It is then an interesting question, both from fundamental and
practical points of view, whether one can witness nonlocality
relying solely on one- and two-body expectation values.
Intuitively, they contain the least information about the
correlations. However, they are also the easiest-to-measure
quantities from which a Bell inequality can be constructed. It has
recently been shown that such Bell inequalities are powerful
enough to witness nonlocality in multipartite quantum states
\cite{our,Lars}. In particular, in Ref. \cite{our} we proposed
classes of permutationally invariant Bell inequalities involving
only two-body correlators that detect nonlocality in multipartite
quantum systems for any number of parties. The main aim of this
paper is to follow the approach of \cite{our}, and search for all
three- and four-partite Bell inequalities that obey a less
restrictive symmetry: translational invariance. We group the Bell
inequalities found within this approach into equivalence classes
under certain symmetries, and check whether they are violated by
quantum states. We then provide an example of a translationally
invariant five-partite Bell inequality, constructed from two-body
correlators, that involve only nearest neighbors, and check that
it is violated by a genuinely multipartite entangled quantum
state. We also show that any translationally invariant Bell
inequality can be maximally violated by a translationally
invariant state, when all parties measure the same set of
observables.

\section{Preliminaries}

We begin by summarizing some known results on Bell inequalities and setting up
the notation we will use throughout the paper.

Consider first $N$ spatially separated parties $A_1,\ldots, A_N$ sharing some
$N$-partite quantum state $\rho$. Each party is allowed to perform measurements
on its share of $\rho$. We restrict the study to the simplest scenario of party
$A_i$ freely choosing between two observables $\mathcal{M}^{(i)}_{x_i}$
($x_i=0,1$), each having two outcomes $a_i=\pm 1$. The correlations that arise
in such an experiment are described by a collection of conditional probabilities
\begin{equation}
\{p(a_1,\ldots,a_N|x_1,\ldots,x_N)\}_{a_1,\ldots,a_N;x_1,\ldots,x_N},
\end{equation}
with $p(a_1,\ldots,a_N|x_1,\ldots,x_N)$ denoting the probability of
obtaining results $a_1,\ldots,a_N$ upon measuring
$\mathcal{M}_{x_1}^{(1)},\ldots,\mathcal{M}_{x_N}^{(N)}$. In our case, i.e.,
when each party chooses between two dichotomic observables, correlations
can be equivalently described by a collection of expectation values
\begin{equation}\label{correlators}
\{\langle \mathcal{M}_{x_{i_1}}^{(i_1)}\ldots
\mathcal{M}^{(i_k)}_{x_{i_k}}\rangle\}_{i_1,\ldots,i_k;x_{i_1},\ldots,x_{i_k};k}
\end{equation}
where $x_{i_1},\ldots,x_{i_k}=0,1$, $i_1<\ldots<i_k=1,\ldots,N$, and
$k=1,\ldots,N$. Notice that the set (\ref{correlators}) contains all mean values
of the local observables and mean values of their products involving up to $N$
parties. It follows that for a Bell experiment with two dichotomic observables
per site, the set (\ref{correlators}) has $3^N-1$ elements, and it is convenient
to think of them, after being ordered, as components of a vector
$\boldsymbol{c}$ belonging to $\mathbbm{R}^{D}$ with $D=3^N-1$. In what follows,
by a slight abuse of terminology, we will identify sets (\ref{correlators}) with
the corresponding vectors. Also, expectation values involving at least two
parties will be called correlators.

Recall that in quantum theory
\begin{equation}
\label{meanvalue} \langle \mathcal{M}_{x_{i_1}}^{(i_1)}\ldots
\mathcal{M}^{(i_k)}_{x_{i_k}}\rangle =\Tr[\rho_{A_{i_1}\ldots
A_{i_k}}(\mathcal{M}_{x_{i_1}}^{(i_1)}\otimes \ldots \otimes
\mathcal{M}^{(i_k)}_{x_{i_k}})],
\end{equation}
where $\rho_{A_{i_1}\ldots A_{i_k}}$ stands for a subsystem of $\rho$
representing the quantum state held by the parties $A_{i_1},\ldots,A_{i_k}$,
i.e., the partial trace of $\rho$ over all the remaining parties and now
$\mathcal{M}_{x_i}^{(i)}$ with $x_i=0,1$ and $i=1,\ldots,N$
denote Hermitian operators with eigenvalues $\pm1$. It is known that the set of
quantum correlations, denoted $\mathcal{Q}$, which arise in the above experiment
when the dimension of $\rho$ is unconstrained, is convex (cf. Ref.
\cite{BrunnerReview}). As a proper subset it contains those correlations that,
even if obtained from quantum states, the parties can simulate by using local
strategies and some shared classical information represented by a random
variable $\lambda$ with probability distribution $p(\lambda)$. As no quantum
resources are needed to create them, we call such correlations
\textit{classical} or \textit{local}. They form a polytope, denoted
$\mathbbm{P}_N$, whose extremal points are those vectors (\ref{correlators}) in
which all correlators factorize, that is,
\begin{equation}\label{factorization}\langle\mathcal{M}_{x_{i_1}}^{(i_1)}
\ldots\mathcal{M}^{(i_k)}_{x_{i_k}}\rangle=
\langle\mathcal{M}_{x_{i_1}}^{(i_1)}\rangle\cdot\ldots\cdot\langle\mathcal{M}^{
(i_k)}_{x_{i_k}}\rangle,
\end{equation}
and each individual mean value $\langle \mathcal{M}_{x_i}^{(i)}\rangle$
($x_i=0,1$, $i=1,\ldots,N$) equals either $-1$ or $1$. In other words, any
vertex $\boldsymbol{v}\in\mathbbm{P}_N$ represents correlations that the parties
can produce by using local \textit{deterministic} strategies, i.e., each local
measurement has a perfectly determined outcome. Denoting then by $V_N$ the set
of vertices of $\mathbbm{P}_N$, one finds that $|V_N|=2^{2N}$, while
$\dim\mathbbm{P}_N=3^N-1$.

Classical correlations are thus represented by those vectors
$\boldsymbol{c}$ that can be written as a convex combination of
vertices $\boldsymbol{v}_{\lambda}\in V_N$,
\begin{equation}
\boldsymbol{c}=\sum_{\lambda}p(\lambda)\boldsymbol{v}_{\lambda},
\end{equation}
where the random variable $\lambda$ with probability distribution
$p(\lambda)$ denotes the shared classical information among the
parties. If $\boldsymbol{c}$ does not admit such a decomposition,
the corresponding correlations are called \textit{nonlocal}. A
multipartite state $\rho$ is then named nonlocal if one can
generate nonlocal correlations from it.

For further purposes it is worth recalling that quantum
correlations are not the only nonlocal ones. Indeed, there exists
a larger set of correlations called nonsignalling that admit this
same feature. It contains those correlations that satisfy the
no-signalling principle: information cannot be transmitted
instantaneously. In terms of a conditional probability
distribution generated in the above experiment, this means that
any of its marginals observed by any group of $N-1$ parties does
not depend on the choice of measurement made by the remaining
party, i.e.,
\begin{eqnarray}\label{nsconstraints}
\sum_{a_i}p(a_1,\ldots,a_i,\ldots,a_N|x_1,\ldots,x_i,\ldots,
x_N)\nonumber\\
=\sum_{a_i}
p(a_1,\ldots,a_i,\ldots,a_N|x_1,\ldots,x'_i,\ldots,x_N)
\end{eqnarray}
for all values of $a_1,\ldots,a_{i-1},a_{i+1},\ldots,a_N$,
$x_1,\ldots,x_{i-1},x_i,x_i',x_{i+1},\ldots,x_N$ and any
$i=1,\ldots,N$. Within the representation we use here to describe
correlations, i.e., in terms of correlators (\ref{correlators}),
this condition is automatically satisfied. In fact, for two-output
measurements labeled by $\pm 1$, any observed conditional
probability distribution can be written as
\begin{eqnarray}
&&p(a_1,\ldots,a_i,\ldots,a_N|x_1,\ldots,x_i,\ldots,
x_N)=\nonumber\\ &&\frac{1}{2^N}\left(1+\sum_{k=1}^{N}\sum_{1\leq
i_1<\ldots<i_k\leq N}a_{i_1}\cdot\ldots \cdot a_{i_k} \langle
\mathcal{M}^{(i_1)}_{x_{i_1}}\ldots \mathcal{M}^{(i_k)}_{x_{i_k}}
\rangle\right).
\end{eqnarray}
It is easy to see that the equalities~(\ref{nsconstraints}) are
satisfied by these conditional probabilities. Thus, any vector
$\boldsymbol{c}\in\mathbbm{R}^D$ represents nonsignalling
correlations if for any choice of $x_1,\ldots,x_N=0,1$ and
$a_1,\ldots,a_N=\pm1$, its elements satisfy the following
inequality
\begin{eqnarray}\label{positivity}
1+\sum_{k=1}^{N}\sum_{1\leq i_1<\ldots<i_k\leq N}a_{i_1}\cdot\ldots \cdot
a_{i_k}
\langle \mathcal{M}^{(i_1)}_{x_{i_1}}\ldots \mathcal{M}^{(i_k)}_{x_{i_k}}
\rangle\geq 0,
\end{eqnarray}
which corresponds to the condition that $p(a_1,\ldots,a_N|x_1,\ldots,x_N)\geq
0$. Geometrically, analogously to the classical correlations, nonsignalling
correlations form a polytope, denoted $\mathcal{N}$, whose facets are given by
inequalities (\ref{positivity}). It follows that
$\mathbbm{P}_N\subset\mathcal{Q}\subset\mathcal{N}$ and it is known that
in general $\mathbbm{P}_N\neq\mathcal{Q}$ \cite{Bell} and $\mathcal{Q}\neq
\mathcal{N}$ \cite{PR}.

The usual tool to detect nonlocality of quantum states are \textit{Bell
inequalities} \cite{Bell}. These are inequalities satisfied by all elements
of $\mathbbm{P}_N$, while possibly being violated by some quantum correlations.
Generally, they can be written as
\begin{eqnarray}\label{BellIneq}
I&:=&\boldsymbol{\alpha}\cdot\boldsymbol{c}\nonumber\\
&=&\sum_{k=1}^N\sum_{1\leq
i_1<\ldots<i_k\leq
N}\sum_{x_{i_1},\ldots,x_{i_k}=0}^{1}\alpha_{x_{i_1},\ldots,x_{i_k}}^{i_1,\ldots
, i_k }
\langle\mathcal{M}^{(i_1)}_{x_{i_1}}\ldots\mathcal{M}^{(i_k)}_{x_{i_k}}\rangle
\geq -\beta_C,\nonumber\\
\end{eqnarray}
where $\alpha_{x_{i_1},\ldots,x_{i_k}}^{i_1,\ldots,i_k}$
are the components of a vector $\boldsymbol{\alpha}\in\mathbbm{R}^D$
and $\beta_C=-\min_{\boldsymbol{c}\in\mathbbm{P}_N}\boldsymbol{\alpha}
\cdot\boldsymbol{c}$ is the so-called classical bound of (\ref{BellIneq}).
Notice that since $\mathbbm{P}_N$ is a polytope, it is enough to optimize $I$
over $\boldsymbol{v}\in V_N$ in order to determine $\beta_C$.
Accordingly, by $\beta_{Q}=-\min_{\boldsymbol{c}\in\mathcal{Q}}I$
and $\beta_{N}=-\min_{\boldsymbol{c}\in\mathcal{N}}I$
we will be denoting the maximal violations of $I$ by quantum and nonsignalling
correlations, respectively. Clearly, $\beta_C\leq \beta_Q\leq \beta_N$ and, in
particular, if $\beta_Q=\beta_C$, a Bell inequality does not have quantum
violation, while if $\beta_N=\beta_C$, it cannot be violated by any
nonsignalling correlations. In the latter case we call a Bell inequality
\textit{trivial}. It is worth mentioning that $\beta_N$ can be efficiently
determined via linear programming and therefore to prove that
a Bell inequality lacks quantum violation it is in some cases easier to show
that $\beta_N=\beta_C$. We will use this fact later in Secs \ref{Sec:N3} and
\ref{Sec44}.

Geometrically, Bell inequalities are half-spaces that constrain $\mathbbm{P}_N$.
In fact, since the latter is a polytope, a finite number of Bell inequalities is
necessary to fully determine it. That is, the set $\mathbbm{P}_N$ is fully
described once all Bell inequalities corresponding to
its facets --- often referred to as \textit{tight} --- are known. This
problem can be fully solved for the simplest scenarios using computer algorithms
such as the cdd algorithm \cite{cdd}; it was, for instance, solved for $N=3$
\cite{Sliwa}, and by imposing the permutational symmetry also for $N=4,5$
\cite{JD}. However, since the dimension of $\mathbbm{P}_N$ and the number of its
vertices grow exponentially with $N$, the problem quickly becomes
computationally intractable for larger $N$.

\section{Translationally invariant two-body Bell inequalities}

\subsection{Bell inequalities from two-body correlators}
\label{Sec31}

We now study the problem of detecting nonlocality in a multipartite
quantum state having access only to one and two-body expectation values
\begin{equation}\label{one_body}
\langle \mathcal{M}_{x_i}^{(i)}\rangle, \qquad
\langle \mathcal{M}_{x_i}^{(i)}\mathcal{M}_{x_j}^{(j)}\rangle
\end{equation}
with $x_i,x_j=0,1$ and $i<j=1,\ldots,N$. As in the general case,
the most straight-\linebreak forward way of tackling this task is to construct
the corresponding polytope of local correlations, denoted $\mathbbm{P}_{N}^{2}$.
This is achieved by getting rid of all correlators of order larger than two in
all elements of $\mathbbm{P}_N$. That is, all correlations $\boldsymbol{c}$
can be written $\boldsymbol{c}=\boldsymbol{c}_{2}\oplus \boldsymbol{c}'$, where
$\boldsymbol{c}_{2}\in\mathbbm{R}^{2N^2}$ and
$\boldsymbol{c}'\in\mathbbm{R}^{D-2N^2}$ stand for parts of
$\boldsymbol{c}$ containing only one and two-body
expectation values, and correlators of order higher than two, respectively.
Denoting by $P:\mathbbm{R}^D\mapsto \mathbbm{R}^{2N^2}$ the
projection $P\boldsymbol{c}=\boldsymbol{c}_{2}$ for any
$\boldsymbol{c}$, one then formally has
that $\mathbbm{P}_{N}^{2}=\{P\boldsymbol{c}\,|\,
\boldsymbol{c}\in\mathbbm{P}_N\}$. In particular, every vertex of
$\mathbbm{P}_N$ is uniquely mapped onto a vertex of $\mathbbm{P}_N^2$, i.e., a
vector $\boldsymbol{v}_{2}$ whose all two-body correlators
factorize ($\langle
\mathcal{M}_{x_i}^{(i)}\mathcal{M}^{(j)}_{x_j}\rangle=\langle
\mathcal{M}_{x_i}^{(i)}\rangle\langle \mathcal{M}^{(j)}_{x_j}\rangle$ for all
$x_i,x_j=0,1$ and $i<j=1,\ldots,N$)
and every one-body expectation value equals
$\pm1$. To see explicitly this one-to-one correspondence, let us denote by
$V_N^2$ the set of extremal elements of $\mathbbm{P}_N^2$ and notice that for
any vertex $\boldsymbol{v}_{2}\in V_N^2$, $\boldsymbol{v}_{2}=P\boldsymbol{v}$
with $\boldsymbol{v}\in V_N$ (cf. \cite{JD} for a proof). On the other
hand, $P\boldsymbol{v}\in V_N^2$ for any
$\boldsymbol{v}\in\mathbbm{P}_N$. Indeed, assume that for some
vertex $\boldsymbol{v}\in
V_N$, $P\boldsymbol{v}=\sum_{i}p_i\boldsymbol{v}_{2}^{i}$, where
$p_i\in(0,1)$ and $\boldsymbol{v}_{2}^{i}$ are some vertices (distinct) of
$\mathbbm{P}_N^2$. This means that at least one of the components of
$P\boldsymbol{v}$ is different than $\pm1$, meaning that
$\boldsymbol{v}$ is not a vertex of $\mathbbm{P}_N$. As a consequence,
$P\boldsymbol{v}\in V_N^2\Leftrightarrow \boldsymbol{v}\in V_N$, that
is vertices of $\mathbbm{P}_N$ are mapped to vertices of $\mathbbm{P}_N^2$, and
any element of $V_N^2$ is a projection of a vertex from $V_N$. It is eventually
straightforward to see that $\boldsymbol{v}_1\neq\boldsymbol{v}_2\Leftrightarrow
P\boldsymbol{v}_1\neq P\boldsymbol{v}_2$ for any
$\boldsymbol{v}_1,\boldsymbol{v}_2\in V_N$. In conclusion
$|V_N^2|=|V_N|=2^{2N}$.

Having characterized the vertices of the polytope $\mathbbm{P}_N^2$, the problem
of witnessing nonlocality from one- and two-body mean values can then be
addressed by determining all its facets. That is, one has
to find all tight Bell inequalities with two-body correlators, which most
generally can be written as
\begin{eqnarray}\label{2corrBI}
I&:=&\sum_{i=1}^{N}(\alpha_i\langle \mathcal{M}^{(i)}_0\rangle
+\beta_i\langle \mathcal{M}^{(i)}_1\rangle)+\sum_{i<j}^{N}\gamma_{ij}
\langle \mathcal{M}_{0}^{(i)}\mathcal{M}_{0}^{(j)}\rangle+\nonumber\\
&&+\sum_{i\neq j}^{N}\omega_{ij}
\langle
\mathcal{M}_{0}^{(i)}\mathcal{M}_{1}^{(j)}\rangle+\sum_{i<j}^{N}\varepsilon_{ij}
\langle \mathcal{M}_{1}^{(i)}\mathcal{M}_{1}^{(j)}\rangle
\geq -\beta_C,
\end{eqnarray}
where $\alpha_i,\beta_i,\gamma_{ij},\omega_{ij}$, and $\epsilon_{ij}$
are some real parameters, and $\beta_{C}$ is the classical bound of $I$:
$\beta_{C}=-\min_{\mathbbm{P}_N^2}I$. In what follows we will shortly call any Bell
inequality of the form (\ref{2corrBI}) a \textit{two-body Bell inequality}.

Although the dimension of the local polytope $\mathbbm{P}_N^{2}$ is
$2N^2$ and, unlike $\dim\mathbbm{P}_N$, grows only polynomially in $N$,
the number of vertices is preserved ($|V_N^2|=|V_N|=2^{2N}$),
meaning that the problem of determining facets of
$\mathbbm{P}_N^2$ remains intractable for larger $N$.
One can, nevertheless, further simplify the classical polytope by
demanding that the Bell inequalities (\ref{2corrBI}) obey some symmetries.
For instance, Ref. \cite{our} focuses on Bell inequalities that are
invariant under any permutation of the parties (see also Ref. \cite{JD} for
symmetric Bell inequalities with full correlators). In that case the dimension
of the local polytope is five irrespective of the
number of parties, and the number of its vertices grows quadratically in $N$. In
this work we consider a less restrictive translational symmetry, i.e., the one
where a Bell inequality is not changed if all parties are shifted by one to the
right modulo $N$. Here, the dimension of the local polytope is not constant with
$N$ as in the permutationally invariant case. However, it is reduced from
quadratic to linear in $N$.

\subsection{Imposing the translational symmetry}
We now move on to imposing translational invariance
to the Bell inequalities (\ref{2corrBI}), that is, we demand that they remain
the same if the transformations
\begin{equation}\label{transformations}
\mathcal{M}^{(i)}_j \rightarrow \mathcal{M}_j^{(i+1)}\qquad (j=0,1),
\end{equation}
are simultaneously applied to all parties with the convention
that $\mathcal{M}_j^{(N+n)}=\mathcal{M}^{(n)}_j$ for any $n=1,\ldots,N$ and
$j=0,1$. This directly translates to certain conditions on the
parameters appearing in (\ref{2corrBI}), namely,
\begin{equation}
\alpha_{i}=\alpha_{i+1}, \qquad \beta_{i}=\beta_{i+1}\qquad (i=1,\ldots,N-1).
\end{equation}
which imply that all $\alpha_i$ and $\beta_i$ must be equal.
Then, $\gamma_{ij}$, $\omega_{ij}$, and $\epsilon_{ij}$ must satisfy the
following cycles of equalities:
\begin{eqnarray}\label{lala}
\gamma_{1,1+k}&=&\gamma_{2,2+k}=\ldots =\gamma_{N-k,N}\nonumber\\
&=&\gamma_{1,N-k+1}=\gamma_{2,N-k+2}=\ldots=\gamma_{k,N}
\end{eqnarray}
and
\begin{eqnarray}\label{lola}
\epsilon_{1,1+k}&=&\epsilon_{2,2+k}=\ldots =\epsilon_{N-k,N}\nonumber\\
&=&\epsilon_{1,N-k+1}=\epsilon_{2,N-k+2}=\ldots=\epsilon_{k,N}
\end{eqnarray}
with $k=1,\ldots,\lfloor N/2\rfloor$, and finally
\begin{eqnarray}\label{lalo}
\omega_{1,1+k}&=&\omega_{2,2+k}=\ldots=\omega_{N-k,N}\nonumber\\
&=&\omega_{N-k+1,1}=\omega_{N-k+2,2}=\ldots=\omega_{N,k}
\end{eqnarray}
with $k=1,\ldots,N-1$.

Denote $\alpha:=\alpha_i$ and $\beta:=\beta_i$ ($i=1,\ldots,N$), and
by $\gamma_k$, $\epsilon_k$ ($k=1,\ldots,\lfloor N/2\rfloor$),
and $\omega_k$ ($k=1,\ldots,N-1$) those parameters that form cycles in
(\ref{lala}), (\ref{lola}), and (\ref{lalo}) enumerated by $k$.
Then, a general translationally invariant Bell inequality with one and
two-body mean values, in what follows referred to as \textit{translationally
invariant two-body Bell inequality}, reads
\begin{eqnarray}\label{BITr}
\alpha\mathcal{S}_0+\beta\mathcal{S}_1+\sum_{k=1}^{\lfloor
N/2\rfloor}\left(\gamma_k
\mathcal{T}_{00}^{(k)}+\epsilon_k\mathcal{T}_{11}^{(k)}\right)
+\sum_{k=1}^{N-1}\omega_k \mathcal{T}_{01}^{(k)}\geq -\beta_C.
\end{eqnarray}
Here, $\mathcal{S}_j$ stand for symmetrized local expectation values, i.e.,
\begin{equation}\label{SymExp}
\mathcal{S}_j=\sum_{m=1}^{N}\langle \mathcal{M}^{(m)}_j\rangle \qquad (j=0,1)
\end{equation}
and $\mathcal{T}_{ij}^{(k)}$ for all translationally invariant two-body
correlators, given explicitly by
\begin{equation}\label{TIExp}
\mathcal{T}^{(k)}_{ij}=\sum_{m=1}^{N}\langle \mathcal{M}_i^{(m)}
\mathcal{M}_j^{(m+k)}\rangle \qquad (i\leq j=0,1),
\end{equation}
with $k=1,\ldots,\lfloor N/2\rfloor$ for $i=j$ and $k=1,\ldots,N-1$ for $i<j$.

Again, the most efficient translationally invariant Bell inequalities
are the facets of the corresponding ``translationally invariant'' local
polytope. To determine the latter, let $P_T$ be a map projecting
every $\boldsymbol{c}_2\in \mathbbm{R}^{2N^2}$ onto a vector consisting of
translationally invariant correlators (\ref{SymExp}) and (\ref{TIExp}), i.e., a
vector of the form
\begin{equation}\label{eq:ti-vertex}
(\mathcal{S}_0,\mathcal{S}_1,\mathcal{T}_{00}^{(1)},
\ldots,\mathcal{T}_{00}^{(\lfloor\frac{N}{2}\rfloor)},
\mathcal{T}_{01}^{(1)},\ldots,\mathcal{T}_{01}^{(N-1)},
\mathcal{T}_{11}^{(1)},\ldots,\mathcal{T}_{11}^{(\lfloor\frac{N}{2}\rfloor)}).
\end{equation}
Then, the polytope of all translationally invariant local correlations is
defined as $\mathbbm{P}_N^{2,T}=P_T\mathbbm{P}_N^2=\{P_T\boldsymbol{c
}_2\,|\,\boldsymbol{c}_2\in\mathbbm{P}_N^2\}$. The dimension of
$\mathbbm{P}_{N}^{2,T}$ is $N+1+2\lfloor N/2\rfloor$, i.e., it grows linearly
with the number of parties $N$. Then, as in the previous case (cf. Sec.
\ref{Sec31}), vertices of $\mathbbm{P}_{N}^{2,T}$, whose set we denote by
$V_N^{2,T}$, must arise as projections of vertices of the polytope
$\mathbbm{P}_N^2$. Precisely, for any $\boldsymbol{v}_{2,T}\in V_N^{2,T}$ there
is $\boldsymbol{v}_2\in V_N^2$ such that
$\boldsymbol{v}_{2,T}=P_T\boldsymbol{v}_2$. Although we are unable to determine
$|V_N^{2,T}|$, it is at most the number of vertices of
$\mathbbm{P}_N^2$ \textit{modulo} translational symmetry (notice that $P_T$ can
map two different vertices of $\mathbbm{P}_N^2$ onto the same vertex of
$\mathbbm{P}_N^{2,T}$), which can be counted by using P\'olya enumeration
theorem \cite{Polya}.

To be more precise, observe that every deterministic local strategy can be
thought of as a function $f: X \longrightarrow Y$, where
$X=\{A_{1},\ldots,A_{N}\}$ is the set of parties and
$Y=\{(1,1),(1,-1),(-1,1),(-1,-1)\}$ is the set of possible pairs of
predetermined outcomes assigned to each party's pair of observables.
Let us now denote by $Y^X$ the set of all such functions, which is isomorphic to
$V_N^2$, and let $G$ be a subgroup of the group of permutations of $N$ parties.
Let us then denote by $Y^X/G$ the set of deterministic local strategies
\textit{modulo} $G$; that is, two strategies $f_1$ and $f_2$ are equivalent if
$f_2$ can be obtained from $f_1$ by permuting the parties with an element of
$G$. Then, P\'olya's enumeration theorem \cite{Polya} states that
\begin{equation}\label{MBR}
\left|Y^X/G\right|=\frac{1}{|G|}\sum_{\sigma \in G}|Y|^{c(\sigma)},
\end{equation}
where the sum runs over all elements of $G$ and $c(\sigma)$ counts the number
of disjoint cycles in the representation of $\sigma$ as a permutation.

For the considered translational symmetry, $G$ is the group
generated by the full cycle $1\mapsto 2 \mapsto \cdots \mapsto N \mapsto 1$,
denoted $\tau$. The group $G$ has $N$ elements of the form
$\tau^k$  ($k=1,\ldots,N$) with $\tau^N$ being the identity element of
$G$. Due to the fact that the number of cycles of $\tau^k$ as a permutation
of an $N$-element set is precisely the greatest common divisor of $N$ and $k$,
denoted $\gcd(N,k)$, Eq. (\ref{MBR}) can be rewritten as
\begin{equation}
\left|Y^X/G\right|=\frac{1}{N}\sum_{k=1}^{N}|Y|^{\gcd(N,k)}
\end{equation}
The above expression can be further simplified by noting
that for any $k$, $d'=\gcd(k,N)$ is in particular a divisor of $N$, i.e.,
$d'|N$, and also by taking into account the fact that some $k$'s have
the same greatest common divisor with $N$. Let then for any divisor $d'$ of
$N$, $\{1\leq k \leq N\,|\, \gcd(N,k)=d'\}$ be the set of all those
$k$'s whose greatest common divisor with $N$ is $d'$. One finds that this set
has the same number of elements as the set
$\{1\leq l \leq N/d'\,|\, \gcd(N/d',l)=1\}$, whose cardinality is, by
definition, the Euler totient function $\varphi(N/d')$. All this means that
\begin{equation}
\left|Y^X/G\right|=\frac{1}{N}\sum_{d'\;\mathrm{s.\,t.\;}d'|N}^{N}
\varphi(N/d')|Y|^ { d' }.
\end{equation}
Using the fact that if $d'|N$, then $d=N/d'$ is
also a divisor of $N$ and substituting $|Y|=4$, we finally arrive at the upper
bound on the number of vertices of $V_N^{2,T}$:
\begin{equation}
\left|V_N^{2,T}\right|\leq
\frac{1}{N}\sum_{d\;\mathrm{s.}\,\mathrm{t.}\;d|N}\varphi(d)4^{N/d}
 \label{eq:polya2}.
\end{equation}
As we shall see in the following sections, the above bound is not always tight:
for $N=3,5$, (\ref{eq:polya2}) is tight indeed, but not for
$N=4$.

To conclude, let us mention that by repeating the arguments used in Ref.
\cite{JD} to prove theorem 1, one can show that if a two-body Bell inequality
is a facet of $\mathbbm{P}_N$, it must also be a facet of $\mathbbm{P}_N^2$.
Analogously, if a two-body translationally invariant Bell inequality is a facet
of $\mathbbm{P}_N^{2}$ then it is a facet of $\mathbbm{P}_N^{2,T}$. The opposite
implications, however, are in general not true. Nevertheless, while it is rather
hard to expect that a two-body Bell inequality can be a facet of
$\mathbbm{P}_N$, some of translationally invariant two-body Bell inequalities
are indeed facets of $\mathbbm{P}_N^{2}$ (see Sec. \ref{Sec44}).

\section{All three- and four-partite translationally invariant
two-body Bell inequalities}
\label{sec4}

In this section we eventually seek all three- and four-partite
translationally invariant two-body Bell inequalities and
characterize their properties. Using the symmetries described in
Sec. \ref{sect:sym}, we group the found Bell inequalities into
equivalence classes that are listed in Secs. \ref{Sec:N3} and
\ref{Sec44}. We also test them with respect to
the possibility of revealing nonlocality in quantum systems and the way in which
we do this is discussed in Sec. \ref{Sect:Qv}. Finally, in Sec. \ref{Sec45} we
discuss the case of $N=5$ and give an example of a nontrivial five-partite
translationally invariant Bell inequality in which two-body correlators involve
only nearest neighbours.

\subsection{Symmetries}
\label{sect:sym}

According to the symmetries inherent to $\mathbbm{P}_N$,
multipartite Bell inequalities can be grouped into equivalence classes.
Naming of parties, observables and outcomes is arbitrary and this is reflected
in the structure of $\mathbbm{P}_N$. In other words, permutation of
any parties, measurements or outcomes in a given Bell inequality results in
another Bell inequality.

Analogously, translationally invariant Bell inequalities
obey certain symmetries and we will use them below to group the facets
of $\mathbbm{P}_3^{2,T}$ and $\mathbbm{P}_4^{2,T}$ into
equivalence classes. These are the following:
\begin{itemize}
\item Renaming of parties in a cyclical way $\mathcal{M}^{(i)}_j\rightarrow
\mathcal{M}^{(i+1)}_j$ for all $i=1,\ldots,N$ and $j=0,1$. By construction, this
symmetry leaves $\mathbb{P}_N^{2,T}$ invariant.

\item Renaming of observables $\mathcal{M}^{(i)}_0
\leftrightarrow
\mathcal{M}^{(i)}_1$ for all parties. On the level of the Bell inequalities
(\ref{BITr}) this corresponds to the changes
$\alpha \leftrightarrow \beta$, $\gamma_k
\leftrightarrow \epsilon_k$ ($k=1,\ldots,\lfloor N/2\rfloor$) and $\omega_k
\leftrightarrow \omega_{N-k}$ ($k=1,\ldots,\lfloor N/2\rfloor$).

\item Renaming of outcomes of the $j$th observable at all sites, i.e.,
$\mathcal{M}^{(i)}_j \leftrightarrow -\mathcal{M}^{(i)}_j$ for all
$i=1,\ldots,N$. In (\ref{BITr}) this symmetry changes $\alpha \leftrightarrow
-\alpha$ and $\omega_k \leftrightarrow
-\omega_k$ ($k=1,\ldots,N-1$) if $j=0$, and $\beta \leftrightarrow -\beta$ and
$\omega_k \leftrightarrow -\omega_k$ ($k=1,\ldots,N-1$) if $j=1$.

\item Renaming of parties $\mathcal{M}^{(i)}_j \leftrightarrow
\mathcal{M}^{(N-i+1)}_j$ for all $i,j$. This symmetry changes in (\ref{BITr})
$\omega_k
\leftrightarrow \omega_{N-k}$ ($k=1,\ldots,\lfloor N/2\rfloor$).
\end{itemize}
It should be stressed that these symmetries form only a proper subset
of all the aforementioned symmetries of the global polytope $\mathbbm{P}_N$ that
are usually used to classify Bell inequalities (cf. Ref. \cite{Sliwa}). This
restricted choice of symmetries is, however, dictated by the fact that in our
classification we want to preserve the translational invariance of Bell
inequalities; a transformation that is not of the above form if applied to
a translationally invariant Bell inequality may result in another Bell
inequality that is no longer translationally invariant.
Notice, however, that the above symmetries preserve translational invariance of
the found Bell inequalities in the most general case, i.e., when the
coefficients in (\ref{BITr}) are unconstrained. However, if some of them are
zero, then translational invariance does not need to be preserved for the
corresponding correlator and one can exploit further symmetries. For example, if
$\alpha = \beta = 0$ and $N$ is even, applying $\mathcal{M}^{(i)}_j
\leftrightarrow -\mathcal{M}^{(i)}_j$ for any even $i$ and any $j$ leads to the
symmetry changes in (\ref{BITr}): $\gamma_k \leftrightarrow -\gamma_k,
\omega_k\leftrightarrow -\omega_k, \epsilon_k \leftrightarrow -\epsilon_k$ for
all odd $k$.

\subsection{Quantum violation}
\label{Sect:Qv}

To search for quantum violation of the Bell inequalities presented in next
sections we use the fact that every multipartite Bell inequality with
two dichotomic measurements per site can be maximally violated
with a multi-qubit state and real one-qubit local measurements
\cite{Masanes,TV}. Consequently, we can assume that
\begin{equation}
\mathcal{M}_{x_i}^{(i)}=\cos
\phi_{x_i}^{(i)}\sigma_z+\sin\phi_{x_i}^{(i)}\sigma_x,
\end{equation}
where $0\leq \phi_{x_i}^{(i)}\leq\pi/2$ for $x_i=0,1$ and $i=1,\ldots,N$, and
$\sigma_x$ and $\sigma_z$ stand for the real Pauli matrices. A given Bell
inequality is then violated by quantum theory if the corresponding Bell operator
(including also the term $\beta_C\mathbbm{1}$ with $\mathbbm{1}$ being the
identity matrix) has a negative eigenvalue. Then, to find its maximal quantum
violation we optimize the lowest eigenvalue of the corresponding Bell operator
over the parameters $\phi_{x_i}^{(i)}$.

Notice that if we further impose that $\phi^{(i)}_{j}=\phi_j$ for all
$i=1,\ldots,N$ and $j=0,1$, then the Bell operator is also translationally
invariant. One could naively think that in such case the Bell inequality is
maximally violated by a pure translationally invariant state. This is generally
not the case (a particular example is inequality $\#25$ presented in Sec.
\ref{Sec44}), however, one can always construct a multi-qubit translationally
invariant mixed state achieving the maximal violation. To be more explicit, let
$\ket{\psi}\in (\mathbbm{C}^2)^{\otimes N}$ be a pure
state maximally violating a given Bell inequality with the same pair of
observables at each site. Let also $V_d$ be a shift operator defined through
\begin{equation}\label{shift}
V_d\ket{\phi_1}\ldots\ket{\phi_N}=\ket{\phi_N}\ket{\phi_1}\ldots\ket{\phi_{N-1}}
\end{equation}
with $\ket{\phi_i}\in\mathbbm{C}^d$ for every $i$. Then, one checks by hand
that the following translationally invariant mixed state
\begin{equation}
\varrho^{TI}=\frac{1}{N}\sum_{k=0}^{N-1}V_2^k \proj{\psi} (V_2^{\dagger})^k
\end{equation}
violates the Bell inequality maximally with the same observables. Later, in Sec.
\ref{Sec5}
we will show that \textit{any} translationally invariant Bell inequality with
$M$ dichotomic observables per site (not necessarily the two-body one)
can be maximally violated by a translationally invariant mixed state and
the same set of observables at all sites.

\subsection{$N=3$}
\label{Sec:N3}

We now consider the case $N=3$. The general formula (\ref{BITr}) reduces to
\begin{eqnarray}
\alpha\mathcal{S}_0+\beta\mathcal{S}_1+\gamma\mathcal{T}_{00}+\epsilon\mathcal{T
}_{11}
+\omega_1\mathcal{T}_{01}^{(1)}+\omega_2\mathcal{T}_{01}^{(2)}+\beta_C\geq 0,
\end{eqnarray}
where we denoted $\gamma_1$ and $\epsilon_1$ by $\gamma$ and $\epsilon$,
respectively, and also skipped the superscripts in
$\mathcal{T}_{00}^{(1)}$ and $\mathcal{T}_{11}^{(1)}$.
The latter, as well as $\mathcal{T}_{01}^{(1)}+\mathcal{T}_{01}^{(2)}$,
are permutationally invariant, and therefore, for $\omega_1=\omega_2$, one
obtains a symmetric Bell inequality (cf. Ref. \cite{our}).

In this case $\dim\mathbbm{P}_3^{2,T}=6$. By using the cdd algorithm \cite{cdd}
solving the convex hull problem, we find that $\mathbbm{P}_3^{2,T}$ has $38$
facets that under the symmetries discussed in the preceding section are grouped
into $6$ classes presented in Table \ref{Tab1}. Moreover, the polytope has $24$
vertices, meaning that in this case the bound (\ref{eq:polya2}) is saturated.

\begin{longtable}{c|ccccccc}
\caption{The list of classes of three-partite translationally invariant
two-body Bell inequalities.}\label{Tab1}\\
\hline
$\#$&$\beta_C$&$\alpha$&$\beta$&$\gamma$&$\omega_1$&$\omega_2$&$\epsilon$
\\
  \hline
  1& 1 &0& 0& 0& 0& 0& 1 \\
2& 3 &0 &0 &0 &1 &-1 &-1 \\
3& 3 &0 &0 &1 &1 &1 &0 \\
4& 3 &1 &1 &0 &1 &0 &0 \\
5& 3 &2 &0 &1 &0 &0 &0 \\
6& 9 &-1 &-3& -1& 1 &2& 3\\
\hline
\end{longtable}

Noticeably, only the last Bell inequality in Table \ref{Tab1}
is violated by quantum states. By construction, it is a facet of
$\mathbbm{P}_N^{2,T}$, but not of
$\mathbbm{P}_{N}^2$. The rest of the classes in Table \ref{Tab1} are trivial
in the sense that they are not violated by any nonsignalling correlations.

The maximal quantum violation of this only nontrivial inequality is
$\beta_Q=10.02$ (while $\beta_N=13$) and it can be realized with
the following pure state
\begin{eqnarray}
\label{fabregas_paquete}
 |\psi_3\rangle &=&
-0.08(|000\rangle+\ket{111})
-0.5628(|001\rangle+|010\rangle+|100\rangle)\nonumber\\
&&+0.1108(|011\rangle
+|110\rangle+|101 \rangle),
\end{eqnarray}
and the measurements given by $\phi_0^{(i)} = -1.1946$ and
$\phi_1^{(i)} = 0.0957$ for all $i=1,2,3$; both chosen so that the
coefficients in front of $\ket{000}$ and $\ket{111}$ are equal. Let us now
shortly discuss the properties of the state $\ket{\psi_3}$. First, it is
symmetric and therefore genuinely multipartite entangled \cite{AnnPhys}. Second,
all its bipartite reductions are local in the sense that they do not violate the
CHSH Bell inequality \cite{chsh}. Then, one notices that the state
$\ket{\psi_3}$ is quite close to the three-qubit $W$ state
$\ket{W_3}=(1/\sqrt{3})(\ket{001}+\ket{010}+\ket{100})$ suggesting that the
latter also violates this Bell inequality. Indeed, one finds that the maximal
violation of our Bell inequality by $\ket{W_3}$ is $9.85$ with
the measurements given by $\phi_0^{(i)} = 5.2556$ and $\phi_1^{(i)} = 0.2285$
$(i=1,2,3)$.

Let us finally check how entangled
is the state $\ket{\psi_3}$ by computing its geometric measure of entanglement.
Recall that for pure $N$-partite states from $(\mathbbm{C}^d)^{\otimes N}$
the latter is defined as \cite{GME}:
\begin{equation}\label{measure}
 E_G(\ket{\varphi})=1-\max_{\ket{\varphi_{\mathrm{prod}}}}|\langle
\varphi_{\mathrm{prod}}|\varphi\rangle|^2,
\end{equation}
where the maximum is taken over all fully product $N$-partite states
$\ket{\varphi_{\mathrm{prod}}}=\ket{e_1}\otimes\ldots\otimes\ket{e_N}
\in(\mathbbm{C}^d)^{ \otimes N}$ with $\ket{e_i}\in\mathbbm{C}^d$ being local
single-party states. By exploiting the fact that for pure symmetric states
the maximum in Eq. (\ref{measure}) is always realized by a symmetric product
vector $\ket{e}^{\otimes N}$ \cite{symmetric}, $E_G$ of our state
$\ket{\psi_3}$ can be computed almost by hand giving
$E_G(\ket{\psi_3})=0.2726$. It is worth pointing out that
$E_G(\ket{W_3})=1/3$ and therefore although the state $\ket{\psi_3}$ is less
entangled than the $W$ state  with respect to $E_G$, it gives a stronger
violation of the above Bell inequality than $\ket{W_3}$.

\subsection{$N=4$}
\label{Sec44}

For $N=4$, the general form of a translationally invariant Bell inequality
(\ref{BITr}) is
\begin{eqnarray}
&&\alpha \mathcal{S}_0+\beta \mathcal{S}_1 +\gamma_1 \mathcal{T}_{00}^{(1)}
+\gamma_2\mathcal{T}_{00}^{(2)}+\epsilon_1
\mathcal{T}_{11}^{(1)}+\epsilon_2\mathcal{T}_{11}^{(2)}\nonumber\\
&&\hspace{2cm}+\omega_1 \mathcal{T}_{01}^{(1)}+\omega_2
\mathcal{T}_{01}^{(2)}+\omega_3\mathcal{T}_{01}^{(3)}+\beta_C\geq 0.
\end{eqnarray}

Here $\dim \mathbbm{P}_{4}^{2,T}=9$,
while the number of different local deterministic strategies amounts to
$|V_4^2|=2^{2\cdot 4}= 256$. \textit{Modulo} translational invariance, this
number reduces to $(1\cdot4^4+1\cdot4^2+2\cdot 4)/4=70$, which, through
(\ref{eq:polya2}), upper bounds the actual number of elements of
$V_4^{2,T}$. By using the cdd algorithm \cite{cdd} we find that $68$ out
of these $70$ ``translationally invariant" local strategies correspond to
extremal vertices of $\mathbbm{P}_{4}^{2,T}$, i.e., the bound (\ref{eq:polya2})
is not tight in this case. This is because the deterministic
local strategies $\{(1,1), (-1,1), (1,-1),(-1,-1)\}$ and $\{(1,1), (-1,-1),
(1,-1),(-1,1)\}$ which are translationally inequivalent give exactly the same
one and two-body translationally invariant correlators, as well as $\{(1,1),
(1,-1), (-1,1),(-1,-1)\}$ and $\{(1,1), (-1,-1), (-1,1),(1,-1)\}$ do. Hence,
they correspond to the same vertex of $\mathbbm{P}_{4}^{2,T}$.

Furthermore, the cdd algorithm \cite{cdd} shows that there are 1038
tight Bell inequalities in this scenario, which we group into 103 classes,
collected in Table \ref{Tab2}. Inequalities \#1 to \#20 are trivial because
$\beta_{N}=\beta_{C}$. Inequalities \#21 to \#24 are not violated by quantum
correlations, although they are violated by nonsignalling correlations.
Noticeably, these inequalities are a bit in the spirit of the
Guess-Your-Neighbour's-Input (GYNI) Bell inequalities \cite{gyni}. That is, they
represent distributed tasks at which quantum theory does not provide any
advantage over classical correlations, while there exist supra-quantum
nonsignalling correlations outperforming them. Recall that GYNI Bell
inequalities are facets of $\mathbbm{P}_N$, whereas our inequalities are not;
still inequality \#21 is a facet of $\mathbbm{P}_N^2$.

Then, all the remaining Bell inequalities (\#25 to \#103) are violated by
quantum correlations. Inequalities \#25 to \#63 can be violated maximally
by using the same pair of measurements at all sites ($\beta_Q=\beta_Q^{TI}$),
and also, except for inequality \#27, they are maximally violated by pure states
that are genuinely multipartite entangled. In particular, inequalities \#26 to
\#63 are maximally violated by translationally invariant pure states, whereas
inequality \#25 by a state orthogonal to the subspace
spanned by translationally invariant states; still it can be maximally violated
by a translationally invariant mixed state (cf. Sec. \ref{Sect:Qv}).

On the other hand, inequalities \#64 to \#103 and inequality \#27 are maximally
violated by states that are product across some bipartition. In particular
inequality \#27 is a sum of two CHSH Bell
inequalities \cite{chsh}: one between $A_1$ and $A_3$ and the other one between
$A_2$ and $A_4$. It is then maximally violated by a product of two two-qubit
maximally entangled states.
Accordingly, in all the cases from \#64 to \#103, different pairs of measurement
settings are required to achieve maximal violation ($\beta_{Q} >
\beta_{Q}^{TI}$). Noticeably, for inequalities \#64 to \#69 the same settings at
all sites do not suffice to achieve quantum
violation ($\beta_{Q}^{TI}=\beta_{C}$).

It should also be stressed that in the case of inequalities \#25, \#73, \#81,
\#87, and \#88, $\beta_Q^{TI}$ is obtained with multipartite states whose all
bipartite subsystems are local. Consequently, nonlocality of these states
cannot be revealed by any bipartite Bell inequality. In this sense, some of the
above violations are purely multipartite, even if obtained from only bipartite
correlations.

Let us finally mention that inequalities 4, 10, 17, 20, 21, 25, 28, 36,
38, 43, 51, 54, 57, 69, 81, 84, 89, 94 are also facets of the two-body local
polytope $\mathbbm{P}_N^2$.


\begin{longtable}{c|c|c|c|rrrrrrrrrr}
\caption{The list of all classes of four-partite translationally
invariant two-body Bell inequalities. As before, by $\beta_{N}$, $\beta_{Q}$,
$\beta_C$ we denote the maximal value of the Bell inequality for nonsignalling,
quantum, and classical correlations, respectively. Then, $\beta_{Q}^{TI}$ stands
for maximal quantum violation with the same observables per site (recall that in
this case the corresponding Bell operator is permutationally
invariant).}\label{Tab2}\\

\hline
 $\#$ & $\beta_{NS}$ & $\beta_Q$ & $\beta_Q^{TI}$ & $\beta_C$ & $\alpha$ &
$\beta$ & $\gamma_1$ & $\omega_1$ & $\omega_3$ & $\epsilon_1$ & $\gamma_2$ &
$\omega_2$ & $\epsilon_2$
 \\
  \hline
\endfirsthead
\caption{continued.}\\
  \hline
$\#$ & $\beta_{NS}$ & $\beta_Q$ & $\beta_Q^{TI}$ &$\beta_C$ & $\alpha$ & $\beta$
& $\gamma_1$ & $\omega_1$ & $\omega_3$ & $\epsilon_1$ & $\gamma_2$ & $\omega_2$
& $\epsilon_2$
\\
  \hline
\endhead
  \multicolumn{6}{l}{{Continued on the next page\ldots}} \\
\endfoot
  \hline
\endlastfoot
1&       4      &4.00&4.00&4&0&2&0&0&0&0&0&0&1\\
2&       4      &4.00&4.00&4&1&1&0&0&1&0&0&0&0\\
3&       4      &4.00&4.00&4&1&1&0&0&0&0&0&1&0\\
4&       4      &4.00&4.00&4&0&-2&0&0&0&2&0&0&1\\
5&       4      &4.00&4.00&4&0&0&0&0&1&1&0&1&0\\
6&       4      &4.00&4.00&4&0&2&0&0&0&1&0&0&0\\
7&       4      &4.00&4.00&4&0&0&0&1&1&0&0&0&1\\
8&       4      &4.00&4.00&4&0&0&0&0&0&-2&0&0&1\\
9&       8      &8.00&8.00&8&-1&-1&-1&1&1&1&1&1&0\\
10&       8      &8.00&8.00&8&-1&1&1&1&1&1&0&1&1\\
11&       8      &8.00&8.00&8&0&2&-1&-1&1&1&0&0&0\\
12&       8      &8.00&8.00&8&1&1&0&1&1&-2&0&-1&1\\
13&       8      &8.00&8.00&8&-1&3&0&-1&-1&2&0&-1&1\\
14&       8      &8.00&8.00&8&-2&-2&1&1&1&1&0&2&0\\
15&       8      &8.00&8.00&8&-2&-2&0&2&2&0&1&0&1\\
16&      16      &16.00&16.00&16&0&4&0&2&2&4&1&2&1\\
17&      16      &16.00&16.00&16&-4&0&2&2&2&2&1&2&1\\
18&      16      &16.00&16.00&16&2&2&-2&3&1&-2&1&-2&1\\
19&      20      &20.00&20.00&20&-3&5&0&-3&-3&4&0&-3&2\\
20&      20      &20.00&20.00&20&-1&5&-1&-2&-2&5&1&-3&1\\
\hline
21&     44/3     &12.00&12.00&12&0&2&-2&0&2&2&1&0&0\\
22&    116/5     &20.00&20.00&20&0&4&-1&-3&3&3&-1&2&0\\
23&      32      &28.00&28.00&28&-2&-8&-2&-2&4&4&-1&2&2\\
24&      32      &28.00&28.00&28&-2&-8&-4&0&4&4&1&0&2\\
\hline
25&     48/5     &8.42&8.42&8&-2&-2&1&0&1&1&0&1&0\\
26&      12      &9.27&9.27&8&0&0&-2&-1&-1&2&1&0&1\\
27&      16      &11.31&11.31&8&0&0&0&0&0&0&-1&2&1\\
28&     76/5     &12.26&12.26&12&0&2&-2&0&0&2&1&2&0\\
29&     76/5     &12.97&12.97&12&0&2&-3&-1&-1&1&1&2&0\\
30&     52/3     &13.60&13.60&12&0&0&-1&-1&-2&4&0&-1&2\\
31&      20      &14.42&14.42&12&0&0&1&1&2&2&-1&3&1\\
32&      20      &14.77&14.77&12&0&0&0&-1&-1&4&-1&-2&2\\
33&     96/5     &16.60&16.60&16&-2&-2&1&2&2&-4&0&0&3\\
34&     96/5     &16.72&16.72&16&-4&-4&1&2&2&0&0&2&1\\
35&     64/3     &17.25&17.25&16&-1&-1&-5&2&2&1&3&-1&0\\
36&      24      &17.50&17.50&16&1&3&1&2&2&3&-1&3&2\\
37&      24      &18.02&18.02&16&0&0&-2&-3&-3&4&1&0&3\\
38&      24      &18.37&18.37&16&-2&-2&2&2&2&2&-1&4&1\\
39&    116/5     &20.36&20.36&20&-2&-4&0&2&2&4&1&4&0\\
40&      24      &20.77&20.77&20&-2&-8&-1&2&2&4&0&0&2\\
41&      24      &20.84&20.84&20&-2&-8&-1&1&1&4&0&2&2\\
42&      28      &21.18&21.18&20&-2&4&0&-2&-2&4&-1&-4&2\\
43&      28      &21.89&21.89&20&0&2&-6&2&2&2&3&-2&0\\
44&      28      &21.93&21.93&20&0&2&-4&-2&-2&0&2&4&-1\\
45&      32      &25.01&25.01&24&0&4&-2&2&4&0&-1&-4&3\\
46&      32      &25.30&25.30&24&-2&-2&-6&4&4&2&5&0&1\\
47&      32      &28.40&28.40&28&-2&-8&-4&0&0&4&1&4&2\\
48&    156/5     &28.41&28.41&28&-6&8&1&-4&-4&4&0&-4&2\\
49&    156/5     &28.48&28.48&28&-6&8&0&-4&-4&4&1&-4&2\\
50&      36      &29.20&29.20&28&-1&-7&-2&4&4&2&0&-3&4\\
51&      36      &29.25&29.25&28&-2&4&-2&-6&-4&6&1&-2&4\\
52&      36      &29.29&29.29&28&-4&-4&0&3&3&4&2&6&-1\\
53&      44      &31.73&31.73&28&-2&4&1&-3&-3&7&-2&-6&3\\
54&      44      &31.84&31.84&28&-2&4&0&-2&-2&6&-2&-6&3\\
55&      40      &33.64&33.64&32&-4&0&0&4&4&-4&-1&-6&3\\
56&    124/3     &36.57&36.57&36&-4&8&-1&-5&-6&8&0&-5&4\\
57&      52      &39.11&39.11&36&-2&4&-4&-8&-8&6&3&0&6\\
58&      60      &42.82&42.82&36&2&-4&2&-4&-4&8&-3&-8&4\\
59&      48      &40.92&40.92&40&-4&8&-2&-8&-6&8&1&-4&5\\
60&      56      &42.32&42.32&40&-4&8&1&-5&-5&9&-2&-8&4\\
61&      64      &50.49&50.49&48&-4&8&-4&-10&-10&8&3&-2&7\\
62&    500/7     &62.89&62.89&60&-14&16&-4&-8&-8&4&5&-4&2\\
63&      104     &74.50&74.50&64&-4&8&4&-8&-8&12&-5&-14&7\\
\hline
64&      10      &8.83&8.00&8&-2&0&1&1&-1&1&0&0&0\\
65&      18      &16.56&16.00&16&-3&-1&1&-2&3&1&0&2&-1\\
66&     56/3     &16.59&16.00&16&2&2&-2&1&-1&-2&1&2&1\\
67&     68/3     &20.46&20.00&20&-2&4&-1&-2&-4&4&0&-2&2\\
68&      24      &21.24&20.00&20&-2&4&-2&-2&-4&4&1&-2&2\\
69&    100/3     &29.15&28.00&28&-2&4&-2&-2&-6&6&0&-2&3\\
70&     44/3     &12.52&12.05&12&-1&3&-2&-1&-2&2&1&0&1\\
71&      24      &21.31&20.06&20&-2&6&1&-3&0&4&-1&-3&1\\
72&    192/5     &32.69&32.10&32&-1&-7&-1&-4&5&5&-2&4&1\\
73&     96/5     &16.45&16.18&16&-2&-2&1&-2&4&1&-1&2&-1\\
74&     44/3     &12.60&12.21&12&-1&-1&-3&1&2&1&2&0&0\\
75&     96/5     &16.60&16.28&16&-4&-2&2&-1&3&1&0&2&-1\\
76&      20      &17.66&16.36&16&0&4&0&-2&2&2&-1&2&1\\
77&      20      &17.66&16.43&16&-1&-5&0&-1&2&3&-1&2&1\\
78&      28      &24.47&24.44&24&-4&8&1&-2&-2&4&-1&-4&2\\
79&      16      &13.66&12.58&12&0&2&1&0&2&2&-1&2&1\\
80&      24      &21.43&20.63&20&-2&-8&0&0&2&4&-1&2&2\\
81&      24      &21.43&20.68&20&-4&2&2&-4&2&2&0&-2&-1\\
82&    500/7     &61.83&60.69&60&-10&12&-12&-8&-8&4&9&0&2\\
83&      32      &25.00&24.70&24&0&4&2&2&4&4&-1&4&3\\
84&      16      &13.66&12.75&12&0&2&-2&2&2&2&2&0&1\\
85&      36      &31.31&28.82&28&-2&8&1&-4&0&6&-2&-4&2\\
86&      44      &36.89&36.85&36&-4&8&0&-5&-5&8&-1&-6&4\\
87&      44      &39.31&36.90&36&-6&-12&1&0&4&4&-2&4&2\\
88&      28      &25.31&24.90&24&-8&-4&3&3&0&1&1&3&-1\\
89&      24      &21.43&20.93&20&-2&-8&-2&2&2&4&1&0&2\\
90&      16      &13.66&12.93&12&0&2&-1&0&2&0&-1&-2&1\\
91&      16      &13.66&13.14&12&-2&-2&1&1&1&0&-1&2&1\\
92&      36      &31.31&29.15&28&-2&-8&-1&-2&2&4&-2&4&2\\
93&      32      &27.31&25.19&24&-2&6&-4&-4&-4&4&3&0&3\\
94&      20      &17.66&17.21&16&0&4&-2&2&2&2&1&-2&1\\
95&      40      &30.62&29.27&28&-2&4&-4&-6&-6&4&3&0&4\\
96&      44      &39.31&37.82&36&-6&-12&-4&4&4&4&3&0&2\\
97&      36      &31.31&29.87&28&-2&-8&-4&4&4&4&2&-2&3\\
98&      44      &38.70&37.98&36&-2&-8&-8&4&4&4&3&-4&2\\
99&      40      &30.85&30.06&28&-2&4&-8&-4&-4&4&5&0&2\\
100&    268/5     &46.34&46.15&44&-10&12&3&-4&-4&4&-2&-8&2\\
101&      32      &23.31&22.77&20&-2&4&1&-2&-2&4&-2&-4&2\\
102&      88      &68.91&67.63&64&-4&8&-8&-14&-14&8&5&2&9\\
103&      80      &69.75&67.97&64&-4&-16&-8&10&10&8&3&-6&7\\
\end{longtable}

\subsection{$N=5$}

\label{Sec45}
Let us finally consider the five-partite case. Here,
$\dim \mathbbm{P}_{5}^{2,T}=10$ and the local polytope $\mathbbm{P}_N^2$ has
$2^{2\cdot 5}=1024$ vertices, which \textit{modulo} translational
invariance reduce to $(1\cdot 4^5+ 4\cdot 4)/5=208$. Interestingly, all the
resulting translationally invariant correlations uniquely correspond to vertices
of $\mathbbm{P}_{5}^{2,T}$, thus, like for $N=3$, the bound (\ref{eq:polya2}) is
saturated in this case. Then, using the algorithm \cite{cdd} one finds that the
polytope $\mathbbm{P}_{5}^{2,T}$ has $34484$ facets, which, after applying the
symmetries from Section \ref{sect:sym} can be grouped into $4198$ different
classes. This number is already too large to list explicitly all these Bell
inequalities. The case $N=6$ is already intractable.

Let us finally mention that the complexity of the local polytope can
further be simplified by imposing additional constraints on the Bell
inequalities. For instance, one can require that a TI Bell inequality contains
only correlators between nearest neighbors. A general form of such an
$N$-partite Bell inequality is
\begin{equation}
\label{eq:5parties}
\alpha \mathcal{S}_0 +\beta\mathcal{S}_1+
\gamma\mathcal{T}_{00}^{(1)}+\omega_1\mathcal{T}_{01}^{(1)}+\omega_2\mathcal{T}_
{01}^{(N-1)}+\epsilon\mathcal{T}_{11}^{(1)}+\beta_C\geq 0
\end{equation}
and the corresponding local polytope has dimension six for any number of
parties. As an illustrative example we consider a five-partite Bell inequality
of this form with $\alpha=-2$, $\beta=-6$, $\gamma=-2$, $\omega_1=2$,
$\omega_2=4$, and $\epsilon=5$. For this choice of parameters one finds that
$\beta_C=35$.

Interestingly, the resulting Bell inequality is capable of
revealing nonlocality in multipartite quantum states. First, assuming
that all parties measure the same pair of single-qubit observables, the maximal
quantum violation of this Bell inequality amounts to $35.29$ and
it is for instance realized by the following five-qubit translationally
invariant state
\begin{eqnarray}
\ket{\psi_5}&=&-0.3710(|00000\rangle+|11111\rangle)
-0.1817 |T_{00001}\rangle
+0.1260 |T_{00011}\rangle\nonumber\\
&&-0.1418 |T_{00101}\rangle
+0.2645 |T_{00111}\rangle
-0.0603 |T_{01011}\rangle\nonumber\\
&&+0.0486 |T_{01111}\rangle,
\end{eqnarray}
where the states $\ket{T_{abcde}}$ are defined through the shift operator
(\ref{shift}) as
\begin{equation}
\ket{T_{abcde}}=\sum_{k=0}^{4}V_2^k\ket{abcde}.
\end{equation}
The corresponding measurements are given by $\phi_0^{(i)} = 1.2967$ and
$\phi_1^{(i)} = 1.9866$ with $i=1,\ldots,5$.
Again, both the state and the measurements are chosen so the coefficients
in front of $\ket{0}^{\otimes 5}$ and $\ket{1}^{\otimes 5}$ are the same.
As one checks, the state $\ket{\psi_5}$ is
genuinely multipartite entangled and all its bipartite subsystems are local in
the sense that they do not violate the CHSH Bell inequality.
Moreover, its geometric measure of entanglement (\ref{measure}) amounts to
$E_G(\ket{\psi_5})=0.4980$.

If one then considers any possible real qubit measurements at all sites,
then the maximal quantum violation of this Bell inequality is
$\beta_Q\approx36.21$. It is, however, realized by the state
$\ket{\psi_5'}=\ket{001}\otimes
(0.7312\ket{00}-0.3775\ket{01}+0.2674\ket{10}+0.5013\ket{11})$ which is
entangled only at the last two qubits. The corresponding measurements are given
by $\phi_0^{(i)} = 0$ for all $i$ and
$\phi_1^{(i)} = 0$ with $i=1,2,3$, and $\phi_{1}^{(4)}=4.7378$ and
$\phi_1^{(5)}=1.2083$.

\section{Bell violation with translationally invariant multiqudit states}
\label{Sec5}

Following the results of Ref. \cite{Moroder}, we now show that the maximal
quantum violation $\beta_Q$ of any translationally invariant Bell inequality
with $M$ dichotomic measurements per site can always be attained with
multipartite quantum states obeying translational symmetry and the same set of
observables at all sites (below referred to as \textit{symmetric} measurement
settings). The local dimension sufficient to do so is $dN$, where $d$ is the
local dimension of a pure state (not necessarily translationally invariant)
violating the Bell inequality maximally. We then provide a numerical method
seeking both the translationally invariant multipartite quantum states and
observables (the same at all sites) that realize $\beta_Q$ for a given Bell
inequality. This algorithm is further tested on Bell inequalities presented in
Table~2, on which it shows good performance. In particular, all these Bell
inequalities are violated by translationally invariant states of local
dimensions smaller than $dN$.

\subsection{The construction}\label{upperbound}

Let us go back to the general Bell inequality (\ref{BellIneq})
and generalize it to the case of $M$ observables at each site, i.e.,
$x_{i}=0,\ldots,M-1$ for all $i$. Assume then that it
is translationally invariant, i.e., elements of $\boldsymbol{\alpha}$ obey the
following equalities
\begin{equation}\label{cycles}
\alpha_{x_{i_1}, \ldots, x_{i_k}}^{i_1,\ldots,i_k}=
\alpha_{x_{i_1}, \ldots, x_{i_k}}^{i_1+1,\ldots,i_k+1}
\end{equation}
for all $i_1<\ldots<i_k=1,\ldots,N$, $x_{i_1},\ldots,x_{i_k}=0,\ldots,M-1$
and $k=1,\ldots,N$, and with the convention that if $i_k=N$, then
\begin{equation}
\alpha_{x_{i_1},
\ldots,x_{i_k}}^{i_1+1,\ldots,i_k+1}=\alpha^{i_1+1,\ldots,N+1}_{x_{i_1},\ldots,
x_ {i_k}}\equiv
\alpha^{1,i_1+1,\ldots,i_{k-1}+1}_{x_{i_k},x_{i_1},\ldots,x_{i_{k-1}}}.
\end{equation}
Assume then that this inequality is violated maximally by a pure
$N$-partite quantum state $\ket{\psi}$ belonging to
$(\mathbbm{C}^d)^{\otimes N}$ and local measurements $\mathcal{M}_{x_i}^{(i)}$.
At each site we extend the Hilbert space to $\mathbbm{C}^d\otimes\mathbbm{C}^N$
and construct the following translationally invariant state acting on
$(\mathbbm{C}^d\otimes\mathbbm{C}^{N})^{\otimes N}$:
\begin{eqnarray}
 \fl\label{rhoTI}
 \varrho^{TI} = \frac{1}{N}\sum_{i=0}^{N-1}
V_d^i\proj{\psi}
(V_d^{\dagger})^i\otimes
V_N^i\proj{0,1,\ldots, N-1}_{A_1\ldots A_N}(V_N^{\dagger})^i,
\end{eqnarray}
where $V_d$ stands for the shift operator defined in (\ref{shift}).
Then, the corresponding dichotomic observables acting on
$\mathbbm{C}^d\otimes\mathbbm{C}^N$ are taken
to be the same at all sites and defined as
\begin{equation}
 \label{meas}
 \widetilde{\mathcal{M}}^{(1)}_{j} = \widetilde{\mathcal{M}}^{(2)}_j = \cdots
= \widetilde{\mathcal{M}}^{(N)}_j = \sum_{i=0}^{N-1}
\mathcal{M}^{(i+1)}_j\otimes \proj{i}
\end{equation}
with $j=0,\ldots,M-1$. Now, exploiting (\ref{cycles}) it is fairly
straightforward to check that the given Bell inequality is maximally violated by
the state (\ref{rhoTI}) and settings (\ref{meas}).

Thus, any translationally invariant Bell inequality can always be maximally
violated by a state that obeys the same symmetry defined on a product Hilbert
space whose local dimension is $dN$. In particular, it follows that
all the TI two-body Bell inequalities found here for $N=3$ and $N=4$ (cf. Sec.
\ref{sec4}) can be maximally violated by TI states of local dimension 6 and 8,
respectively.

Although the above construction is general, it remains unclear whether $dN$ is
the minimal local dimension necessary to produce $\beta_Q$ with TI states and
symmetric settings. For instance, in the three-partite case the Bell inequality
$\#6$ is maximally violated by a three-qubit translationally invariant state and
the same pair of measurements at all sites (see Sec. \ref{Sec:N3}). Below we
show that all those Bell inequalities in Table \ref{Tab2} that have quantum
violation are violated by TI states of local dimension lower than eight.

\subsection{Numerical technique}\label{numtech}

Here we discuss an algorithm providing a lower bound on the maximal violation
$\beta_Q$ of a translationally invariant Bell inequality in fixed local
dimensions $D$, assuming that the underlying quantum state is translationally
invariant and the measurement settings are the same at all sites. It also
provides an upper bound on the minimal dimension $d_{\min}$ of the local Hilbert
space required to attain $\beta_Q$ with translationally invariant states and
symmetric settings. In the case of small number of parties (say, $N<6$) and
low dimensions ($D<7$), we conjecture that the algorithm finds
the maximal quantum violation for a given local dimension $D$.
This would imply that the resulting upper bounds on $d_{\min}$ are the best
possible ones, that is, the algorithm finds $d_{\min}$ with high confidence.
In particular, $d_{\min}$'s found by the algorithm
for the Bell inequalities from Table (\ref{Tab2}) improve the general
upper bound $dN$ discussed in Sec. (\ref{upperbound}).

Let us now move on to the algorithm. It is a variant of the see-saw type
iteration technique introduced in Ref. \cite{seesaw1} (see also
Ref. \cite{seesaw2}) and it is formulated as
follows:

\begin{enumerate}
  \item For a given translationally invariant Bell inequality with $M$
dichotomic measurements per site represented by the vector
$\boldsymbol{\alpha}$ [cf. Ineq. (\ref{BellIneq})] fix the dimension of the
local Hilbert space as $D$ for all parties. Then, generate random unitary
matrices $U_j$ $(j=0,\ldots,M-1)$ according to a uniform distribution (see
\cite{randomunitary} for a simple method allowing to do so) and take the first
party's dichotomic observables as
\begin{equation}
\mathcal{M}_j^{(1)}=U_j\Lambda U_j^{\dagger},
\end{equation}
where $\Lambda$ is a $D$-dimensional diagonal matrix with randomly
generated $\pm 1$ entries.

\item Identify the observables of the remaining $N-1$ parties with the
first party's ones, i.e.,
    \begin{equation}
    \mathcal{M}_j^{(i)}=\mathcal{M}_j^{(1)}
    \end{equation}
for all $i=2,\ldots,N$ and $j=0,\ldots,M-1$.

\item Form the corresponding Bell operator
\begin{eqnarray}\label{BellOp}
B=\sum_{k=1}^{N}\sum_{1\leq i_1<\ldots<i_k\leq
N}\sum_{x_{i_1},\ldots,x_{i_k}=0}^{M-1}\alpha^{i_{1},\ldots,i_k}_{x_{i_1},\ldots
,x_{i_k}}\mathcal{M}^{(i_1)}_{x_{i_1}}\otimes\ldots\otimes
\mathcal{M}^{(i_k)}_{x_{i_k}}\nonumber\\
\end{eqnarray}
and minimize $\mathrm{Tr}(B\rho)$ subject to $\rho\geq 0$ and $\Tr\rho=1$. This
amounts to finding the minimal eigenvalue of $B$ with the corresponding
eigenvector $\ket{\psi}$, and the Bell inequality violation
$\beta=-\langle\psi|B|\psi\rangle$.

  \item Generate the translationally invariant state
  \begin{equation}\label{TIstateII}
  \varrho^{TI}=\frac{1}{N}\sum_{i=0}^{N-1}V_D^i\proj{\psi}(V^{\dagger}_D)^i,
  \end{equation}
  where $V_D$ is defined as in Eq. (\ref{shift}) with $d$ replaced by $D$.
Notice that
$\varrho^{TI}$ features
the same violation of the
Bell inequality as
$\ket{\psi}$, i.e., $\Tr(\varrho^{TI}B)=\langle\psi|B|\psi\rangle=-\beta$.

  \item Then, for the Bell operator (\ref{BellOp}) and the
state
(\ref{TIstateII}) one has
\begin{equation}
\Tr(B\varrho^{TI})=\sum_{j=0}^{M-1}\Tr(\mathcal{M}^{(1)}_{j}F_{j}),
\end{equation}
where (to get the expression below we have used the fact that the state
$\varrho^{TI}$ is translationally invariant)
\begin{eqnarray}
F_{j}&=&\sum_{k=1}^N\sum_{2 \leq i_1 < \ldots < i_k \leq N}
\sum_{x_{i_2},\ldots,x_{i_k}=0}^{M-1}
\alpha_{j,x_{i_2}\ldots x_{i_k}}^{1,i_{2}-i_1+1\ldots, i_{k}-i_1+1}
\nonumber\\
&&\times\Tr_{A_{2}\cdots
A_{N}}\left(\mathbbm{1}_{A_1}\otimes
\mathcal{M}_{x_{i_2}}^{(i_2-i_1+1)}\otimes\ldots\otimes
\mathcal{M}_{x_{i_k}}^{(i_{k}-i_1+1)} \varrho^{TI}\right).\nonumber\\
\end{eqnarray}

  Now, minimize the expression
  \begin{equation}
 \widetilde{ \beta}=\sum_{j=0}^{M-1}{\Tr(\mathcal{M}_{j}^{(1)} F_{j})}
  \end{equation}
over all Hermitian matrices $\mathcal{M}_{j}^{(1)}$ such that
$-\mathbbm{1}\leq \mathcal{M}_{j}^{(1)}\leq\mathbbm{1}$ $(j=0,\ldots,M-1)$.
This can easily be done by introducing the eigenvalue decomposition of
each $F_j$,
\begin{equation}
F_j=\sum_{i}\lambda_i^{(j)}\proj{\phi_i^{(j)}}.
\end{equation}
Then, one directly finds that
the optimal first party's measurements are given by
\begin{equation}
\mathcal{M}_j^{(1)}=-\sum_{i}{\mathrm{sgn}(\lambda_i^{(j)})\proj{\phi_i^{(j)}}}
\qquad (j=0,\ldots,M-1),
\end{equation}
and the resulting Bell violation $\widetilde{\beta}\geq \beta$ (however, now the
measurement settings are no longer symmetric).

  \item Go back to step (ii) and keep repeating the protocol until convergence
of the objective value $\beta$ is reached.
\end{enumerate}
The above algorithm differs from the
one proposed in Ref. \cite{seesaw2} in that it contains two
additional steps (ii) and (iv). The first one is to guarantee that the violation
is obtained with the same pair of measurements at each site, while the second
one to ensure that the state realizing this violation is translationally
invariant.
It should be noticed, however, that due to step (ii) it is unclear whether the
violations $\beta$ produced by our version of the algorithm are non-decreasing
at each iteration step. This is because after requiring at step (ii)
that all parties measure the same observables, the violation could in principle
drop. However, our numerical studies show that this is not the case and at each
iteration step of the algorithm the value of $\beta$ is non-decreasing. We then
conjecture that this is always the case.

\subsection{Observations}

We applied the above procedure to inequalities $\#64-\#103$ in Table \ref{Tab2},
i.e., those for which $\beta_Q>\beta_Q^{TI}$ with four-qubit states, leading to
the following observations regarding values of $d_{\min}$:
\begin{itemize}
   \item $d_{\min}\le 6$ for each inequality. That is,
a six-dimensional Hilbert space at each site is enough to obtain
$\beta_Q$ with translationally invariant states in all cases. In
fact, three-dimensional component spaces suffice in the majority of
cases; the only exceptions are:
   \item $d_{\min}\le 4$ for inequalities $\#64$, $\#65$, $\#73$, $\#78$,
$\#81$, $\#86$,
$\#91$, $\#99$, $\#100$,
   \item $d_{\min}\le 5$ for inequalities $\#70,\#82$,
   \item $d_{\min}\le 6$ for inequalities $\#67, \#68, \#69, \#74, \#75$.
 \end{itemize}
We conjecture that the above bounds are all tight, that is, we can
replace the inequality with equality. Moreover, we found that
in all the cases real-valued measurements suffice to saturate
these bounds.

\section{Conclusion}

In this work we have explored the question if Bell inequalities
involving only one and two-body expectation values are capable of
witnessing nonlocality in multipartite quantum states, pursuing
the research started in Ref. \cite{our}. To simplify our
considerations we have considered only those Bell inequalities
that obey the translational symmetry. We have found all tight Bell
inequalities (facets of the corresponding polytopes of classical
correlations) for the three and four-partite cases and grouped
them into equivalence classes under certain symmetries. We have
then characterized their properties and checked whether they are
violated by quantum theory. Noticeably, in some cases the states
violating these inequalities have all bipartite subsystems local,
meaning that their nonlocality cannot be revealed by any bipartite
Bell inequality. We have then provided an example of a
five-partite translationally invariant Bell inequality which
contains correlators involving only the nearest neighbours and
checked that it is powerful enough to detect nonlocality in
multipartite states. Finally, we have shown that any
translationally invariant Bell inequality with $M$ dichotomic
observables per site can always be violated maximally by a
translationally invariant state (mixed in general) and the same
set of observables per site. We have also discussed an algorithm
that finds a maximal quantum violation in the above setting for a
fixed local dimension.

Clearly, our studies can be further developed. For instance, one could
generalize our results to an arbitrary number of parties. The other possibility
is to consider other symmetries than permutational or translational, and also
more complicated scenarios, i.e., with more measurements and more outcomes per
site, leading in both cases to stronger Bell inequalities. A bit more
nontrivial direction is to construct a nontrivial $N$-partite Bell
inequalities consisting of correlators involving only the nearest
neighbours. Such Bell inequalities, being in the spirit of entanglement
witnesses constructed from two-body Hamiltonians \cite{Geza}, would
facilitate nonlocality detection in many-body systems in which correlations
between the nearest neighbours are the dominating ones.

\ack{Discussions with J. Stasi\'{n}ska are greatly acknowledged. This work is supported by Hungarian National Research Fund OTKA (PD101461), Spanish DIQIP CHIST-ERA, TOQATA (FIS2008-00784) and COQPIC (FIS2010-14830)  projects and AP2009-1174 FPU PhD grant, EU IP SIQS, ERC AdG QUAGATUA and OSYRIS, StG PERCENT. This publication was made possible through the support of a grant from the John Templeton Foundation. R. A.
acknowledges the Spanish MINECO for the Juan de la Cierva scholarship.}

\end{document}